\documentclass[aps,prd,nofootinbib,twocolumn,longbibliography]{revtex4-1}
\usepackage[english]{babel}
\usepackage[autostyle, english = british]{csquotes}
\usepackage[CJKbookmarks, pdftex, bookmarksnumbered, bookmarksopen, colorlinks, citecolor=blue, linkcolor=blue]{hyperref}
\usepackage{amsmath,amssymb}
\usepackage{graphicx}
\usepackage{enumitem}
\usepackage{color}

\setlist[enumerate]{topsep=0pt,parsep=-1mm,leftmargin=5mm,}
 \MakeOuterQuote{"}
\def\be{\begin{equation}}
\def\ee{\end{equation}}

\newcommand*{\diff}{\mathop{}\!\mathrm{d}}

\begin{document}

\title{\large Painlev\'e-Gullstrand coordinates discontinuity\\ in the quantum Oppenheimer-Snyder model}

\author{Francesco Fazzini${}^{a}$, Carlo Rovelli${}^{bcd}$ and Farshid Soltani${}^{e}$}

\affiliation{${}^a$ Department of Mathematics and Statistics,
University of New Brunswick, Fredericton,
New Brunswick E3B 5A3, Canada}
\affiliation{${}^b$ Aix-Marseille University, Universit\'e de Toulon, CPT-CNRS, F-13288 Marseille, France.}
\affiliation{${}^c$ Department of Philosophy and The Rotman Institute of Philosophy, 1151 Richmond Street~N, London, Ontario N6A5B7, Canada}
\affiliation{${}^d$ Perimeter Institute, 31 Caroline Street N, Waterloo, Ontario, N2L2Y5, Canada}
\affiliation{${}^e$Department of Physics and Astronomy, University of Western Ontario, London, Ontario N6A 3K7, Canada}

\begin{abstract} 
\noindent A metric that describes a collapsing star and the surrounding black hole geometry accounting for quantum gravity effects has been derived independently by different research groups. There is consensus regarding this metric up until the star reaches its minimum radius, but there is disagreement about what happens past this event. The discrepancy stems from the appearance of a discontinuity in the Hamiltonian evolution of the metric components in Painlev\'e-Gullstrand coordinates.  Here we show that the {\em continuous} geometry that describes this phenomenon is represented by a {\em discontinuous} metric when written in these coordinates.  The discontinuity disappears by changing coordinates. The discontinuity found in the Hamiltonian approach can therefore be interpreted as a coordinate effect. The geometry continues regularly into an expanding white hole phase, without the occurrence of a shock wave caused by a physical discontinuity.

\end{abstract}


\maketitle

The Einstein equations predict a collapsing star and the surrounding spacetime to evolve into regions of very high curvature. Here we expect quantum gravitational effects to come into play, radically modifying the classical dynamics. Effective metrics that take these quantum effects into account have been extensively explored in the last years. Among the most promising is a quantum modification of the Oppenheimer-Snyder (OS) model~\cite{Oppenheimer:1939ue}: a simple description of the gravitational collapse of a homogeneous and pressureless star. The model has been derived  using physical inputs form Loop Quantum Gravity, both canonical and covariant, and borrowing techniques from Loop Quantum Cosmology \cite{Kelly:2020uwj,Kelly:2020lec,Lewandowski:2022zce,Han:2023wxg}. In it, the star reaches a maximum density and a minimum radius, and then bounces.  

There is a remarkable agreement in the literature on the description of the quantum effects on the collapsing phase. But there is some disagreement about what happens when the star reaches its minimum radius.  In the Hamiltonian approach developed in \cite{Kelly:2020uwj,Kelly:2020lec}, which uses generalized Painlev\'e-Gullstrand coordinates, a discontinuity in the metric coefficients develops at the bounce.  This has been interpreted as indicating the onset of a shock wave in the dynamics of gravity, absent in the {continuous} geometry studied in \cite{Lewandowski:2022zce,Han:2023wxg}. Here we show that in this {\em continuous} geometry the Painlev\'e-Gullstrand coordinates---and hence the metric written in these coordinates---become discontinuous at the bounce. This shows that the discontinuity found in the Hamiltonian approach is a coordinate effect.

In what follows, we first recall the classical OS model.  Then we describe its quantum-corrected version recently studied in \cite{Lewandowski:2022zce,Han:2023wxg}.  We then show explicitly how, at the moment of the bounce, a discontinuity is formed in the metric components when it is expressed in generalized Painlev\'e-Gullstrand coordinates.

The OS model describes the gravitational collapse of a homogeneous and pressureless star, and it is the prototypical example of black hole formation by gravitational collapse. In Planck units ($c=G=\hbar=1$), and assuming the star's boundary to be in free fall and to start at rest at past infinity, the metric in the interior of the star reads
\be
\diff s^2 = - \diff T^2 + a^2 (T) \big( \diff R^2 + R^2 \diff \Omega^2\big)\, .
\label{int}
\ee
The coordinate $T$ is the proper time of observers moving at constant radial and angular coordinates, $a(T)$ is the scale factor that determines the size of the star at time $T$, $R\in [0,R_{\text{star}}]$ is the comoving radial coordinate and $\diff \Omega^2$ is the line element of a unit two-sphere. The trajectory of the boundary of the star in the interior metric is given by $R=R_{\text{star}}$. The Einstein field equations give the Friedmann equation for the scale factor:
\be
\frac{\dot{a}^2}{a^2}=\frac{8 \pi}{3} \rho\,,
\label{Fried}
\ee
where the overdot means differentiation with respect to $T$ and
\be
\rho=\frac{m}{\frac{4}{3}\pi \big( R_{\text{star}} a \big)^3}
\ee
is the uniform density of the star. Eq.~\eqref{Fried} can be solved:
\be
a(T)= \bigg(\frac{9m (T-T_0)^2}{2 R^3_{\text{star}}} \bigg)^{1/3}  .
\label{aclassic}
\ee
Without loss of generality we can take the time at which the star collapses to zero physical radius to be $T=0$.

The exterior of the star is described by the Schwarzschild metric
\be
\diff s^2 = - f(r)\diff t^2 + f^{-1}(r)  \diff r^2 + r^2 \diff \Omega^2\, ,
\ee
where
\be
f(r)=1-\frac{2m}{r}\,.
\ee
The worldsheet of the boundary of the star in the exterior spacetime is given by the trajectory $r(T)$, where $T$,  the time coordinate of the interior metric, is also the proper time on the boundary of the star. 

A continuous and differentiable geometry requires the geometry of the interior and the geometry of the exterior to match on the boundary of the star. That is, the induced metric and the extrinsic curvature of the boundary of the star have to match on the two sides. This requirement determines the trajectory of the boundary of the star in the exterior region. 

The metric described above takes a particularly simple form in Painlev\'e-Gullstrand (PG) coordinates, where the matching conditions are manifestly satisfied. The full spacetime metric can be written in a unique coordinate patch using these coordinates.  Let us show this explicitly. Changing the comoving radial coordinate $R$ in the star interior to the area (or Schwarzschild) radius
\be
r(T,R)= a(T) R
\label{rPGin}
\ee
and changing the exterior Schwarzschild time coordinate $t$ to the PG time coordinate
\be
T(t,r)= t + 2\sqrt{2mr} +2m \ln \bigg| \frac{\sqrt{r/2m}-1}{\sqrt{r/2m}+1}\bigg|
\label{TPGc}
\ee
in the exterior region, the full spacetime metric can be written as
\be
\diff s^2 = - \diff T^2 +  \Big( \diff r +  N^r (T,r) \diff T\Big)^2 + r^2 \diff \Omega^2\, ,
\label{OSc}
\ee
where
\be
 N^r (T,r) =    
 \begin{cases}
      - 2r/3T &\quad r\leq r_{\text{star}} (T)\vspace{2mm} \\ 
       \sqrt{2m/r} &\quad r> r_{\text{star}} (T)\\
     \end{cases}
\label{Nc}
\ee
and
\be
r_{\text{star}} (T) = \bigg(\frac{9m T^2}{2} \bigg)^{1/3} \, .
\ee

The same result can be obtained from a Hamiltonian formalism for Lema\^itre-Tolman-Bondi (LTB) spacetimes, of which the Oppenheimer-Snyder model is a particular case. The choice of PG coordinates leads to a particularly simple set of Hamiltonian equations of motion.  Assuming initial conditions corresponding to a constant density dust star and a vacuum exterior, the solution for the whole spacetime is identical to what is obtained from the matching procedure described above  \cite{Kelly:2020uwj}.

Let us next describe the quantum gravitational corrections to this model. Consider first the interior metric in Eq.~\eqref{int}. Loop quantum gravity modifies the Friedmann equation~\cite{Ashtekar:2006rx,Ashtekar:2006uz,Kelly:2020lec} for the scale factor in Eq.~\eqref{Fried} to
\be
\frac{\dot{a}^2}{a^2}=\frac{8 \pi}{3} \rho\, \Big( 1-\frac{\rho}{\rho_c}\Big)\,,
\label{Friedq}
\ee
with the critical density $\rho_c$ being a parameter of Planckian value. This equation can be integrated to give
\be
a(T)= \bigg(\frac{9m T^2 +A m}{2 R^3_{\text{star}}} \bigg)^{1/3},
\label{aq}
\ee
where $A=3/(2\pi\rho_c)$ is a parameter of Planckian value with the dimensions of a squared mass. Eq.~\eqref{aq} gives the quantum correction to the classical equation \eqref{aclassic}. It shows that in the quantum-corrected effective metric, the physical radius of the star never collapses to zero, but it rather reaches its minimum size
\be
r_M=a(0)R_{\text{star}}=(Am/2)^{1/3}
\ee
at $T=0$, before bouncing and starting to increase. 

Let us next address the quantum correction to the exterior metric. This can be derived in two distinct ways, which remarkably converge. It can be simply derived by requiring the exterior geometry to have the same symmetries as in the classical case, and then imposing the matching with the quantum-corrected metric of the star \cite{Lewandowski:2022zce}.  Equivalently, the quantum corrections to the exterior metric coming from loop quantum gravity can be studied separately from the interior region, using an effective Hamiltonian formalism containing LQG corrections adapted to spherical symmetry \cite{Gambini:2020nsf,Kelly:2020uwj, Kelly:2020lec}, checking then the matching at the boundary. These two different roads lead to the same exterior metric. The same Hamiltonian approach can be performed for LTB spacetimes, and for the case that there is an exterior vacuum region, the metric in that region is again the same \cite{Husain:2021ojz}.

Let us write this metric explicitly. Consider a generic spherically symmetric metric having a hypersurface-orthogonal Killing vector field. Its line element reads
\be
\diff s^2 = - F(r)\diff t^2 + G(r)  \diff r^2 + r^2 \diff \Omega^2\, ,
\label{extq}
\ee
where $F$ and $G$ are functions of the radial coordinate and the hypersurface-orthogonal Killing vector is $\partial_t$. In order for the matching with the interior metric to be possible, the exterior parametric trajectory $(t(T),r(T))$ of the boundary of the star needs to be a radial geodesic of the exterior metric. Such trajectories satisfy 
\be
F \,\dot{t} = E\, ,
\ee
where $E$ is the conserved energy associated to $\partial_t$, and
\be
G \,\dot{r}^2 = E/F -1\, .
\ee
The matching of the induced metric and of the extrinsic curvature of the boundary of the star on the two sides uniquely fixes the functions $F$ and $G$ to~\cite{Lewandowski:2022zce}
\be
F(r)=G^{-1}(r)=1-\frac{2m}{r}+\frac{A m^2}{r^4}
\label{F}
\ee
for all $r > r_M$. 

\begin{figure}[t]
	\centering
\includegraphics[scale=.45]{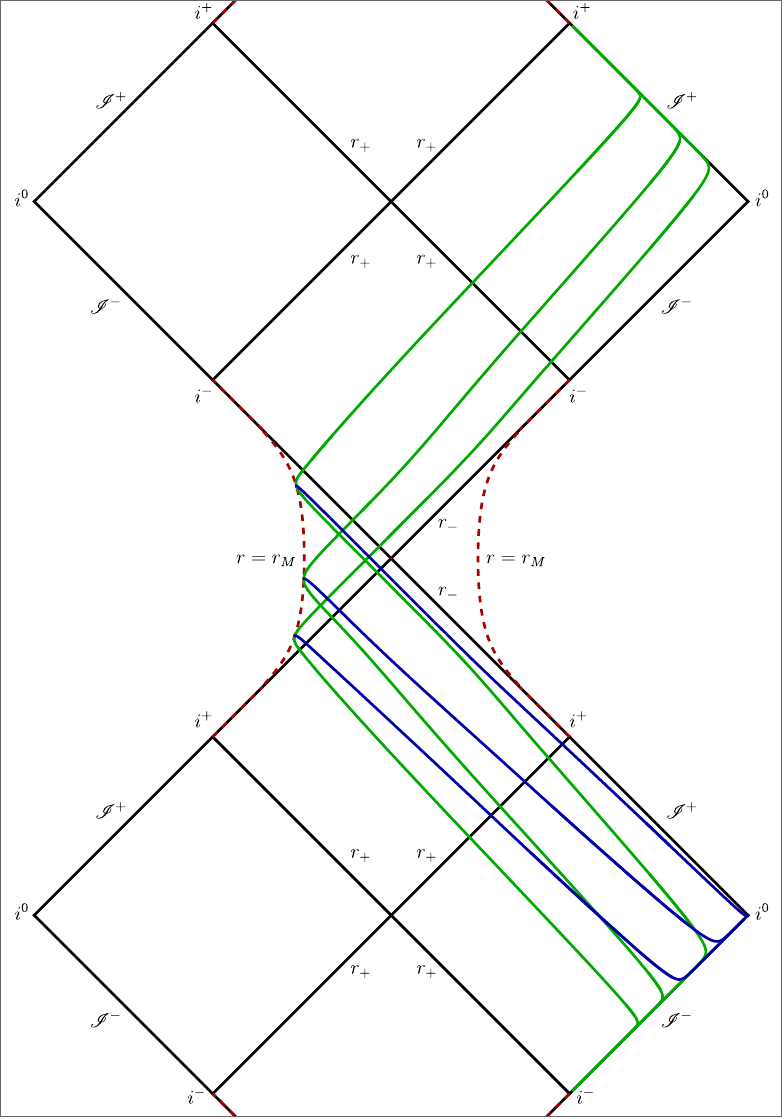}
\caption{\em Section of the conformal diagram of the spacetime defined by Eqs.~\eqref{extq} and \eqref{F} plotted using global Kruskal coordinates defined in~\cite{soltaniGlobalKruskalCoordinates2023} and $m=1$, $A=0.99$ (Planck units). The trajectories of three different radially free-falling observers are plotted in green and the constant $T$ surfaces at the time of their turning point are plotted in blue.}
\label{uno}
\end{figure}

\begin{figure}[b]
	\centering
\includegraphics[scale=.55]{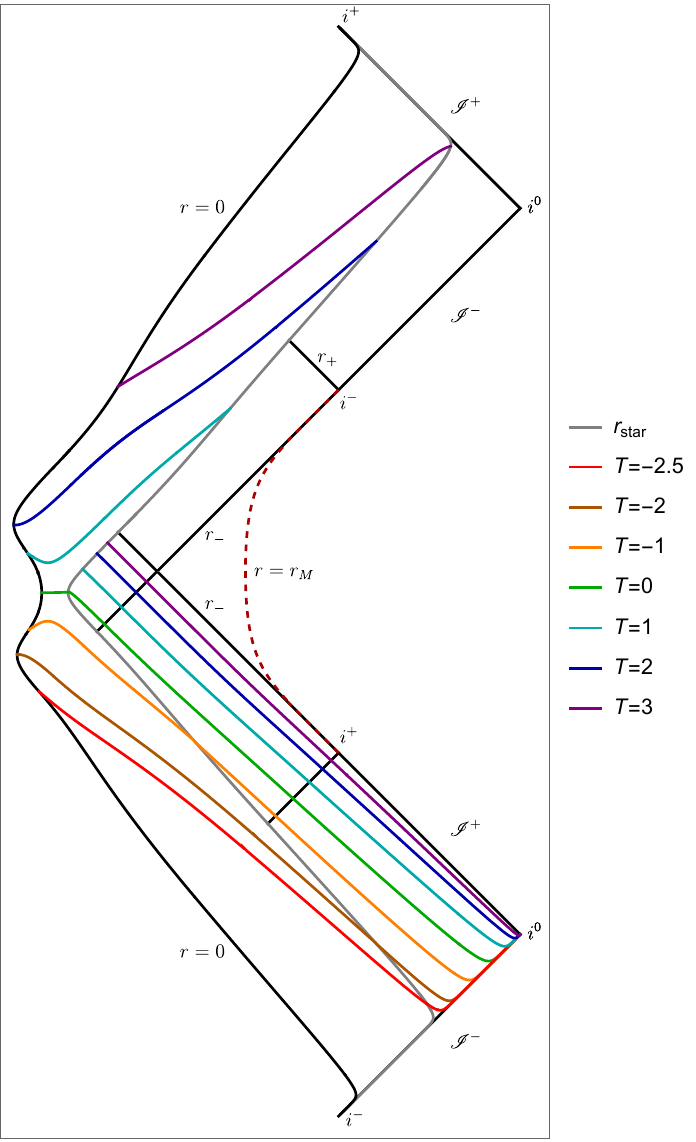}
\caption{\em Conformal diagram of the quantum-corrected OS spacetime with $m=1$ and $A=0.99$ (Planck units).}
\label{due}
\end{figure}

The maximal extension of this exterior geometry in the absence of the star has been extensively studied in~\cite{Lewandowski:2022zce,Munch:2021oqn} and its conformal diagram is reported in Fig.~\ref{uno}. It is very similar to the Reissner-Nordstr\"om geometry. A radially free-falling geodesic, e.g. the wordline of the star's boundary, bounces in an interior region which is neither trapped nor antitrapped, and that separates a trapped (black hole) and an antitrapped (while hole) region.  The conformal diagram of the quantum-corrected OS spacetime is shown in Fig.~\ref{due}. This completes the construction of the quantum-corrected OS model. A further quantum correction of this metric that gets rid of the residual curvature singularity and of the Cauchy horizon, and that has a single asymptotic region, is constructed in~\cite{Han:2023wxg}. 

Now we come to our main observation.  Given the simplicity of the classical OS metric in PG coordinates, it is natural to study also the quantum-corrected geometry in the PG coordinates. Let us see what happens.  The interior metric can be easily written in PG coordinates by performing the coordinate transformation in Eq.~\eqref{rPGin}, where $a(T)$ is now the quantum-corrected scale factor. For the exterior metric, the differential of the PG coordinate time $T$ satisfies the relation
\be
\diff T = \diff t - \frac{\sqrt{1-F}}{F} \diff r\,.
\label{dTPG}
\ee
The full spacetime metric can then be written as
\be
\diff s^2 = - \diff T^2 +  \Big( \diff r +  N^r (T,r) \diff T\Big)^2 + r^2 \diff \Omega^2\, ,
\label{OSq}
\ee
where
\be
 N^r (T,r) =    
 \begin{cases}
      -\frac{6rT}{9T^2 + A} &\quad r\leq r_{\text{star}} (T)\vspace{2mm} \\ 
       \sqrt{1-F(r)} &\quad r> r_{\text{star}} (T)\\
     \end{cases}
\label{Nq}
\ee
and
\be
r_{\text{star}} (T) = \bigg(\frac{9m T^2 + Am}{2} \bigg)^{1/3} \, .
\ee
The metric in Eqs.~(\ref{OSq}) and (\ref{Nq}) is exactly the metric found in~\cite{Kelly:2020uwj,Kelly:2020lec} by separately studying the quantum corrections to the interior and the exterior metric in PG coordinates. This proves the equivalence of the two different constructions for the exterior metric and the overall consistency of the quantum-corrected OS model.

There is a problem however. The function $N^r (T,r)$ in Eq.~\eqref{Nq} becomes discontinuous for $T>0$. This is immediately seen from the fact that the expression valid in the interior of the star changes sign after the bounce while the expression valid outside the star does not.  In~\cite{Kelly:2020lec,Husain:2021ojz} this discontinuity was tentatively interpreted as a physical discontinuity of the gravitational field and it was argued that a shock wave must form as a consequence of it.

We have however now all the ingredients to clarify the reason for the discontinuity of the function $N^r (T,r)$. Consider in fact the interior geometry of the star. The coordinate $T$ is the proper time of observers moving at constant comoving radial coordinate $R$, which in the area coordinate $r$ [see Eq.~\eqref{rPGin}] means
\be
r(T) \propto a(T) \,.
\ee
This means that the time coordinate $T$ in the interior region is well adapted to bouncing observers, with $T=\mathrm{const.}$ surfaces adapted to infalling observers for $T<0$ and to outgoing observers for $T>0$. In the exterior region however, as extensively discussed in Appendix A and shown in Fig.~\ref{uno}, the PG time coordinate $T$ is only adapted to infalling observers, thus creating the discontinuity at the star's surface for $T>0$. This discontinuity in the PG time coordinate can be clearly seen from Fig.~\ref{due}, where $T=\mathrm{const.}$ surfaces for the complete quantum-corrected OS spacetime are plotted.


In~\cite{Kelly:2020lec,Husain:2021ojz} a dust field was used as a relational clock to gauge fix the Hamiltonian constraint.  The time Painlev\'e-Gullstrand coordinate $T$ has then a physical interpretation as a clock time. The chrono-geometry {\em measured by this dust field taken as a clock} is indeed discontinuous.  That is, if one uses surfaces of constant dust-time to reconstruct the 4-dimensional geometry, then a discontinuity in the metric is unavoidable. The above analysis shows that it is possible to interpret this discontinuity as a discontinuity of the clock field, evolving in a continuous geometry.  The geometry in this spacetime is continuous and the discontinuity in the metric comes only from using the dust field as a relational clock. No shock wave is formed.

An interesting feature of the model studied in \cite{Kelly:2020uwj,Kelly:2020lec} was that the star bounces out in the same asymptotic region where it collapsed, with a predicted lifetime of the order of $m^2$.  As shown in \cite{Han:2023wxg}, it is still possible to have a single asymptotic region also when the geometry is continuous during and after the bounce by gluing together different spacetime patches and breaking the local Killing symmetry also around the horizon.\footnote{See also \cite{Munch:2021oqn} where a continuous OS spacetime within the model studied in \cite{Kelly:2020uwj,Kelly:2020lec} is constructed. This however has a global structure very different from the one of the spacetime in Fig.~\ref{due} and it features an infinite number of asymptotic regions.} In this case the lifetime is a free parameter of the spacetime, ultimately to be determined by a quantum gravity calculation.

Confusing coordinate artifacts are common in general relativity. A famous example is the $r=2m$ singularity in the Schwarzschild metric, that prompted Einstein and many others to believe that spacetime ends at the horizon.

\begin{acknowledgments}
We thank Ed Wilson-Ewing for a very extensive collaboration in the development of this work. 
F.S.'s work at Western University is supported by the Natural Science and Engineering Council of Canada (NSERC) through the Discovery Grant "Loop Quantum Gravity: from Computation to Phenomenology". Western University is located on the traditional lands of Anishinaabek, Haudenosaunee, L\=unaap\`eewak, and Attawandaron peoples.
C.R. acknowledges support by the Perimeter Institute for Theoretical Physics  through its distinguished research chair program. Research at Perimeter Institute is supported by the Government of Canada through
Industry Canada and by the Province of Ontario through the Ministry of Economic Development and Innovation.
This work was made possible through the support of the ID\# 62312 grant from the John Templeton Foundation, as part of \href{https://www.templeton.org/grant/the-quantuminformation-structure-ofspacetime-qiss-second-phase}{‘The Quantum Information Structure of Spacetime’ Project (QISS)}. The opinions expressed in this work are those of the authors and do not necessarily reflect the views of the John Templeton Foundation.
F.F.'s work at UNB is supported in part by the Natural Sciences and Engineering Research Council of Canada.%
\end{acknowledgments}

\appendix

\section{Incompleteness of the PG coordinates in the exterior vacuum region}

Consider the exterior vacuum region defined by Eqs.~\eqref{extq} and \eqref{F} in the absence of the star. The conformal diagram of this spacetime for $r>r_M$ is given in Fig.~\ref{uno}. The PG coordinate time $T$ is the proper time of radially free-falling observers that start at rest at infinity. In Schwarzschild spacetime such observers all hit the singularity inside the black hole. But this is no longer the case in the metric in Eqs.~\eqref{extq} and \eqref{F}. In the latter the trajectories of free-falling observers starting at rest at infinity satisfy
\be
\dot{r}^2 = 1-F\, .
\ee
All these trajectories have a turning point at $\dot{r}=\sqrt{1-F}=0$, which is solved by $r=r_M$.  This means that instead of hitting the singularity in the interior of the black hole, radially free-falling observers reach a minimum distance from it at $r=r_M$ and then bounce back out of it in a second future asymptotic region. A few of these trajectories, together with the constant $T$ surfaces at the time of their turning point, are plotted in Fig.~\ref{uno}

From this discussion it is clear that the PG coordinates do not cover the vacuum spacetime region at $r<r_M$. The PG coordinate time $T$ is the proper time of the radially free-falling observers and none of these observers penetrates inside the $r=r_M$ surface: there is no PG coordinate time $T$ inside the region at $r<r_M$. This is consistent with the expression of the metric in PG coordinates given in Eqs.~(\ref{OSq}-\ref{Nq}). In fact, since $N^r (T,r)=\sqrt{1-F(r)}$, the metric is well defined only for $1-F(r)>0$, which is 
\be
r>r_M.
\label{constraint}
\ee
This same constraint was found in~\cite{Kelly:2020uwj} during the construction of the quantum-corrected vacuum exterior region from loop quantum gravity first principles. This constraint was interpreted as a physical property of the quantum-corrected spacetime. The discussion above shows that it is  an artifact of the specific coordinate system employed.   

In fact, this is a general phenomenon. As pointed out in \cite{Faraoni:2020ehi}, the PG coordinates fail wherever the Misner-Sharp mass is negative, as is precisely the case in the metric defined by Eqs.~\eqref{extq} and \eqref{F} for $r<r_M$.

Furthermore, Fig.~\ref{uno} clearly shows that the PG coordinates do not cover the full spacetime region traversed by the radially free-falling observers, and hence also by the boundary of the star in the quantum-corrected OS model, but only the spacetime region they cover before their turning point. There is then no hope for the PG coordinates to provide a global coordinate patch for the quantum-corrected OS model, as instead they do in the classical case.

\bibliographystyle{utcaps}
\bibliography{qOS}

\end{document}